\documentclass[preprint,review,12pt]{elsarticle}
\usepackage[margin=3cm]{geometry}

\usepackage{amssymb, amsmath}

\journal{Nuclear Physics A}

\begin{document}

\begin{frontmatter}



\title{$\Lambda_c N$ interaction from lattice QCD\\and its application to $\Lambda_c$ hypernuclei}


\author[YITP,RIKEN]{Takaya~Miyamoto}
\author[YITP,RIKEN,CCS]{Sinya~Aoki}
\author[RIKEN,iTHEMS]{Takumi~Doi}
\author[RIKEN]{Shinya~Gongyo}
\author[RIKEN,iTHEMS]{Tetsuo~Hatsuda}
\author[RIKEN,RCNP]{Yoichi~Ikeda}
\author[RIKEN,NihonU]{Takashi~Inoue}
\author[RIKEN]{Takumi~Iritani}
\author[RIKEN,RCNP]{Noriyoshi~Ishii}
\author[RIGAKU]{Daisuke~Kawai}
\author[RIKEN,RCNP]{Keiko~Murano}
\author[RIKEN,RCNP]{Hidekatsu~Nemura}
\author[YITP,RIKEN]{Kenji~Sasaki}

\address[YITP]{Center for Gravitational Physics, Yukawa Institute for Theoretical Physics, Kyoto University, Kyoto 606-8502, Japan}
\address[RIKEN]{Theoretical Research Division, Nishina Center, RIKEN, Wako 351-0198, Japan}
\address[CCS]{Center for Computational Sciences, University of Tsukuba, Ibaraki 305-8571, Japan}
\address[iTHEMS]{iTHEMS Program and iTHES Research Group, RIKEN, Wako 351-0198, Japan}
\address[RCNP]{Research Center for Nuclear Physics (RCNP), Osaka University, Osaka 567-0047, Japan}
\address[NihonU]{Nihon University, College of Bioresource Sciences, Kanagawa 252-0880, Japan}
\address[RIGAKU]{Department of Physics, Kyoto University, Kyoto 606-8502, Japan}

\begin{abstract}
The interaction between $\Lambda_c$ and a nucleon ($N$) is investigated by employing the HAL QCD method in the (2+1)-flavor lattice QCD on a $(2.9~\mathrm{fm})^3$ volume at $m_\pi \simeq 410,~570,~700$ MeV. 
We study the central potential in $^1S_0$ channel as well as central and tensor potentials in $^3S_1 - $$^3D_1$ channel, and find that the tensor potential for $\Lambda_c N$ is negligibly weak and central potentials in both $^1S_0$ and $^3S_1 - $$^3D_1$ channels are almost identical with each other except at short distances. 
Phase shifts and scattering lengths calculated with these potentials show that the interaction of $\Lambda_c N$ system is attractive and has a similar strength in $^1S_0$ and $^3S_1$ channels at low energies (i.e. the kinetic energy less than about $40$ MeV). 
While the attractions are not strong enough to form two-body bound states, our results lead to a possibility to form $\Lambda_c$ hypernuclei for sufficiently large atomic numbers ($A$). 
To demonstrate this, we derive a single-folding potential for $\Lambda_c$ hypernuclei from the $\Lambda_c$-nucleon potential obtained in lattice QCD, and find that $\Lambda_c$ hypernuclei can exist for $A \ge 12$ with the binding energies of a few MeV.
We also estimate the Coulomb effect for the $\Lambda_c$ hypernuclei.
\end{abstract}


\end{frontmatter}


\section{Introduction}
The investigation of baryon-baryon interactions is one of the most important subjects to understand properties of hadronic matter. 
The low-energy nucleon-nucleon ($NN$) interaction has been severely constrained by the $NN$ scattering data and the properties of finite nuclei \cite{Machleidt:2001rw}. 
The hyperon-nucleon ($YN$) and hyperon-hyperon ($YY$) interactions have also been investigated phenomenologically to reproduce the properties of hypernuclei and hyperon-nucleon scattering data \cite{Hiyama:2010zzc}.
Such phenomenological interactions are then used to study yet unknown nuclei and also the neutron star interiors.

As a natural extension, it is interesting to investigate charmed hypernuclei, as initiated in Ref.~\cite{Dover} just after the discovery of the $\Lambda_c$ baryon. 
Including charm quarks, the one-boson-exchange potential (OBEP) model for the $Y_c N~(Y_c = \Lambda_c, \Sigma_c)$ was constructed \cite{Dover_OBEP}, where the couplings are determined by assuming the flavor $SU(4)$, which is an extension of the flavor $SU(3)$ for the $YN$ interaction. 
The possibility of both $\Lambda_c$ and $\Sigma_c$ nuclear bound states was predicted for heavy nuclei \cite{Dover}. 
Further studies were carried out in \cite{Bando1, Bando, Bando2}: 
Based on flavor $SU(4)$ symmetry, the authors made a comparison between the $\Lambda$ hypernuclei and the $\Lambda_c$ hypernuclei. 
Although the depth of the effective potential for $\Lambda_c$ in the G-matrix calculation is about $2/3$ of that for $\Lambda$, they found that the number of bound states in $\Lambda_c$ hypernuclei is larger than that in $\Lambda$ hypernuclei due to its heavy mass. 

However, the charm quark is much heavier than other three quarks (up, down, strange), so that the flavor $SU(4)$ may not give a good description of the $Y_c N$ interaction. 
Therefore, Ref.~\cite{Oka} has investigated $\Lambda_c N$ interaction with the OBEP model based on the heavy quark effective theory, where the $\Lambda_c N - \Sigma_c N - \Sigma_c^* N$ coupled channel system is considered. 
It was then found that $\Lambda_c N$ interaction is strong enough to form a 2-body bound state due to effects of these channel couplings, but the results are rather sensitive to the model parameters. 
Studies within a framework of the constituent quark model \cite{Huang,Gal,Maeda:2015hxa,Valcarce} have also been attempted to extract the $\Lambda_c N$ interaction.
In Ref.~\cite{Tsushima1, Tsushima2}, the authors have estimated the single particle energy for $\Lambda_c$ baryon in several nuclei by using the quark-meson coupling model, and they claim that $\Lambda_c$ hypernuclei are likely to be formed. 

The aim of the present paper is to shed a new light on the problem of the $\Lambda_c N$ interaction from first-principles lattice QCD simulations. 
For this purpose, the HAL QCD method to analyze the hadron-hadron interactions on the lattice \cite{HAL1,HAL2,HAL3} is most suitable.
The method has already been applied to various hadron-hadron systems \cite{Inoue1, HAL_CCP, Doi, Murano, Ikeda, Sasaki} and to hadronic matter \cite{Inoue2}. 
Advantages of this method in the context of the present paper are as follows:
(i) Applications to the charmed systems are straightforward,  
(ii) no phenomenological parameters are involved since it is based on first-principles QCD simulations, and 
(iii) the resultant $\Lambda_c N$ potential is faithful to the QCD S-matrix below the inelastic threshold, so that we can correctly calculate the $\Lambda_c N$ elastic scattering phase shift. 

This paper is organized as follows. 
In section \ref{sec:HAL_method}, we present a brief description of the HAL QCD method for the $\Lambda_c N$ system. 
The numerical setup for this work is summarized in section \ref{sec:setup}. 
In section \ref{sec:results_pot}, we show our numerical results of the $\Lambda_c N$ potentials in both $^1S_0$ and $^3S_1-$$^3D_1$ channels. 
We also discuss properties of the $\Lambda_c N$ interaction through phase shifts and scattering lengths calculated by our potentials. 
In section \ref{sec:folding_pot}, we employ the single-folding potential to investigate $\Lambda_c$ hypernuclei.
Summary and conclusions are presented in section \ref{sec:summary}.

\section{HAL QCD method for $\Lambda_c N$ system} \label{sec:HAL_method}
In this section, we discuss the HAL QCD method \cite{HAL1} to be applied to the $\Lambda_c N$ system.
We start with the equal-time Nambu-Bethe-Salpeter (NBS) wave function in the center-of-mass (CM) frame of two baryons at Euclidean time $t$;
\begin{eqnarray}
	\psi^{(W)}_{\alpha \beta} (\vec{r}) e^{-W t} = \sum_{\vec{x}} \langle 0| B^{(1)}_\alpha (\vec{r}+\vec{x}, t) 
	B^{(2)}_\beta (\vec{x}, t) | B^{(1)} (\vec{k}) B^{(2)} (-\vec{k}), W \rangle, \label{eq:NBS_QCD}
\end{eqnarray}
where $\alpha, \beta$ are the spinor indices, and $B_\alpha (\vec{x}, t)$ denotes a local interpolating operator for a baryon $B$. 
The state $| B^{(1)} (\vec{k}) B^{(2)} (-\vec{k}), W \rangle$ stands for an energy eigenstate of a two baryon system. 
Here the total energy is denoted by  $W = \sqrt{|\vec{k}|^2 + m_{B^{(1)}}^2} + \sqrt{|\vec{k}|^2 + m_{B^{(2)}}^2}$ with the baryon masses $m_{B^{(1)}}$ and $m_{B^{(2)}}$, and the relative momentum $\vec{k}$.
We employ the local interpolating operators for a nucleon and $\Lambda_c$ as
\begin{equation}
	B_\alpha (x) = \epsilon_{ijk} \left[ q^T_i (x) C \gamma_5 q_j (x) \right] q_{k, \alpha} (x) \label{eq:Bop_oct}
\end{equation}
where $x = (\vec{x}, t)$, and $i, j, k$ are color indices. $C$ is the charge conjugation matrix defined by $C = \gamma_2 \gamma_4$, and  $q =u,d,c$ stands for quark operators  for up-, down- and charm-quarks, respectively. 
Flavor structures of a nucleon and $\Lambda_c$ are given by
\begin{eqnarray}
	N &\equiv&
	\begin{pmatrix}
		p \\
		n
	\end{pmatrix}
	=
	\begin{pmatrix}
		 \left[ u d \right] u \\
		 \left[ u d \right] d
	\end{pmatrix}, \\
	\Lambda_c &=& \frac{1}{\sqrt{6}} \left( \left[ c d \right] u +  \left[ u c \right] d - 2 \left[ d u \right] c \right).
\end{eqnarray}
In the asymptotic region ($r = |\vec{r}| \to \infty$), the NBS wave function satisfies the Helmholtz equation $\left[ |\vec{k}|^2 + \vec{\nabla}^2 \right] \psi^{(W)}_{\alpha \beta} (r) \simeq 0$ and its asymptotic behavior for a given orbital angular momentum $L$ and total spin $S$ is denoted as
\begin{eqnarray}
	\psi^{(W)}_{LS} (r) \propto e^{i \delta_{LS} (k)} \frac{\sin{(kr - L\pi/2 + \delta_{LS} (k))}}{kr},
\end{eqnarray}
where the ``scattering phase shift" $\delta_{LS} (k)$ is determined from the unitarity of the S-matrix in QCD \cite{HAL2,HAL3}. 
From the NBS wave function, the potential which reproduces the scattering phase shift is defined through the Schr\"{o}dinger equation as
\begin{equation}
	\left( E - H_0 \right) \psi^{(W)}_{\alpha \beta} (\vec{r}) = 
	\int d^3 r^\prime U_{\alpha \beta; \alpha^\prime \beta^\prime} (\vec{r}, \vec{r^\prime}) 
	\psi^{(W)}_{\alpha^\prime \beta^\prime} (\vec{r^\prime}), \label{eq:Sch_eq_NBS}
\end{equation}
where $H_0 = -\vec{\nabla}^2 / 2\mu$ with the reduced mass $\mu = m_{B^{(1)}} m_{B^{(2)}} / (m_{B^{(1)}} + m_{B^{(2)}})$, and $E = k^2 / 2\mu$ is a kinetic energy of the two baryon system in the CM frame.
In this definition, the non-local potential $U (\vec{r}, \vec{r^\prime})$ is energy-independent below the inelastic threshold \cite{HAL2,HAL3}. 
In order to handle the non-locality of the potential, we introduce the derivative expansion as \cite{OME}
\begin{equation}
	U (\vec{r}, \vec{r^\prime}) = V (\vec{r}, \vec{\nabla})~\delta^{(3)} (\vec{r} - \vec{r^\prime}), 
		\label{eq:Derivative_exp}
\end{equation}
where $V (\vec{r}, \vec{\nabla})$ is then expanded in terms of $\vec\nabla$.
For example, the leading order of the derivative expansion is given by
\begin{eqnarray}
	V_{LO} (\vec{r}) &=& V_0 (\vec{r}) + V_\sigma (\vec{r}) (\vec{\sigma_1} \cdot \vec{\sigma_2}) 
	+ V_T (\vec{r}) S_{12}, \nonumber \\
	S_{12} &=& 3 \frac{(\vec{r} \cdot \vec{\sigma_1}) (\vec{r} \cdot \vec{\sigma_2})}{|\vec{r}|^2}
	- (\vec{\sigma_1} \cdot \vec{\sigma_2}),
\end{eqnarray}
where $\vec{\sigma_i}$ is the Pauli matrix acting on the spin index of the $i$-th baryon.
The local potentials $V_0$, $V_\sigma$ and $V_T$, which give the spin-independent force, the spin-spin force and the tensor force, are commonly used in nuclear physics. 
The convergence of the derivative expansion can be checked e.g. by changing the energy.
For example, the leading order approximation is found to be accurate 
for $E < 45$ MeV in the case of the $NN$ scattering in quenched QCD with $m_\pi\simeq 530$ MeV \cite{Edep_NN}.

In lattice QCD, the NBS wave functions can be extracted from the baryon four-point correlation function given by
\begin{equation}
	G_{\alpha \beta} (\vec{r}, t-t_0) = \sum_{\vec{x}} \langle 0| B^{(1)}_\alpha (\vec{r}+\vec{x}, t) 
	B^{(2)}_\beta (\vec{x}, t) \overline{\mathcal{J}^{(J^P)}} (t_0)| 0 \rangle, \label{eq:4pt-corr}
\end{equation}
where $\overline{\mathcal{J}^{(J^P)}}(t_0)$ is the source operator which creates two baryon states with the total angular momentum $J$ and the parity $P$.
Inserting a complete set between the two-baryon operator and the source operator in the Eq.~(\ref{eq:4pt-corr}), we obtain 
\begin{eqnarray}
	G_{\alpha \beta} (\vec{r}, t-t_0) &=& \sum_n \sum_{\vec{x}} \langle 0| B^{(1)}_\alpha (\vec{r}+\vec{x}, t) 
	B^{(2)}_\beta (\vec{x}, t) | W_n \rangle 
	\langle  W_n | \overline{\mathcal{J}^{(J^P)}} (t_0)| 0 \rangle  + \cdots 
	\nonumber \\
	& = &
	 \sum_n \psi^{(W_n)}_{\alpha \beta} (\vec{r}) e^{-W_n (t-t_0)} A_n + \cdots ,
	\label{eq:extract_NBS_lattice}
\end{eqnarray}
with constant $A_n = \langle W_n | \overline{\mathcal{J}^{(J^P)}}(0) | 0 \rangle$, where $| W_n \rangle$ stands for an elastic scattering state with the energy of $W_n$, and the ellipses represent contributions from inelastic states. 
\subsection{Source operator} \label{sec:HAL_method:src_op}
In this work, we choose a wall source at $t=t_0$ defined by
\begin{eqnarray}
	\mathcal{J}^{(J^P) \mathrm{wall}} (t_0) = P^{(J^P)}_{\beta \alpha} \left[ B^{(1) \mathrm{wall}}_\alpha (t_0) 
	B^{(2) \mathrm{wall}}_\beta (t_0) \right], \label{eq:wall_source}
\end{eqnarray}
where $P^{(J^P)}_{\beta \alpha}$ is the projection operator to the total angular momentum $J$ and the parity $P$. 
Here $B^{\mathrm{wall}} (t_0)$ is obtained by replacing the local quark operator $q (\vec{x}, t)$ in the $B (\vec{x}, t)$ with the wall quark operator given by
\begin{eqnarray}
	q^\mathrm{wall} (t_0) \equiv \sum_{\vec{x}} q (\vec{x}, t_0),
\end{eqnarray}
with the Coulomb gauge fixing at $t=t_0$. 
Since the orbital angular momentum of the wall source is fixed to $L=0$, the source with fixed total angular momentum are obtained by the spin projection of the source.
In the case of the $\Lambda_c N$ system, which has the total spin $S=0$ or $S=1$, the wall source operator $\mathcal{J}^{(J^P) \mathrm{wall}}_{\Lambda_c N} (t_0)$ with the spin projection to $S=0$ or $S=1$ creates the $\Lambda_c N$ system with $J^P=0^+$ or $J^P=1^+$, respectively.
\subsection{Time dependent HAL QCD method} \label{sec:HAL_method:time_dep}
In principle, the baryon four-point correlation function is dominated by the NBS wave function of the ground state in the large time separation (Eq.~(\ref{eq:extract_NBS_lattice})). 
In practice, however, it is difficult to realize the ground state domination since $t-t_0$ cannot be taken large enough due to statistical noises of the baryon four-point correlation function \cite{mirage,LAT16sa,LAT16ti,sanity}.
This difficulty was overcome by the time-dependent HAL QCD method \cite{TdepHAL} as follows.
Let us consider the normalized baryon four-point correlation function from Eq.~(\ref{eq:4pt-corr}) as
\begin{eqnarray}
	R_{\alpha \beta} (\vec{r}, t-t_0) &\equiv& 
	\frac{G_{\alpha \beta} (\vec{r}, t-t_0)}{e^{-m_{B^{(1)}}(t-t_0)} e^{-m_{B^{(2)}}(t-t_0)}} \nonumber \\
	&=& \sum_n \psi^{(W_n)}_{\alpha \beta} (\vec{r}) e^{-\Delta W_n (t-t_0)} A_n + \cdots, \label{eq:Rcorr}
\end{eqnarray}
where $\Delta W_n = W_n - (m_{B^{(1)}} + m_{B^{(2)}})$, which satisfies 
\begin{equation}
	E_n \equiv{ k_n^2 \over 2\mu}= \Delta W_n + \frac{1 + 3 \delta^2}{8 \mu} \left( \Delta W_n \right)^2 
	+ \mathcal{O} \left( \left( \Delta W_n \right)^3 \right), \label{eq:relation_E_Wn}
\end{equation}
where $\delta = (m_{B^{(1)}} - m_{B^{(2)}}) / (m_{B^{(1)}} + m_{B^{(2)}})$.
Using the above relation and the Schr\"{o}dinger equation in Eq.~(\ref{eq:Sch_eq_NBS}), we obtain 
\begin{eqnarray}
	\left[ \left( \frac{1 + 3 \delta^2}{8 \mu} \right) \frac{\partial^2}{\partial t^2} -\frac{\partial}{\partial t} - H_0 \right] 
	R_{\alpha \beta} (\vec{r}, t-t_0) = 
	\int d^3 r^\prime U_{\alpha \beta; \alpha^\prime \beta^\prime} (\vec{r}, \vec{r^\prime}) 
	R_{\alpha^\prime \beta^\prime} (\vec{r^\prime}, t-t_0), \label{eq:time-dep_HAL}
\end{eqnarray}
for a moderately large $t-t_0$, where contributions from the inelastic states can be neglected. 
Although the higher order terms in Eq.~(\ref{eq:relation_E_Wn}) can be calculated by corresponding time derivative, those contributions turn out to be numerically negligible in the present lattice setup.
The effects of higher derivative terms and the contribution from inelastic states are regarded as the systematic errors and estimated by the time dependence of scattering observables.
It is noted that Eq.~(\ref{eq:time-dep_HAL}) becomes exact for $m_{B^{(1)}} = m_{B^{(2)}}$. 

\section{Lattice setup} \label{sec:setup}
For numerical simulations, we employ the (2+1)-flavor full QCD configurations generated by PACS-CS Collaboration \cite{PACS-CS} with the renormalization-group improved Iwasaki gluon action and a nonperturbatively $\mathcal{O}(a)$ improved Wilson-clover quark action at $\beta = 6/g^2 = 1.90$ on a $L^3 \times T = 32^3 \times 64$ lattice. 
The corresponding lattice spacing is $a = 0.0907(13)$ fm and physical lattice size is $La = 2.902(42)$ fm. 

In order to see the quark mass dependence of the potentials, we employ three ensembles of gauge configurations. 
The hopping parameters of these ensembles are $\kappa_{ud} = 0.13700$ (Ensemble 1), $0.13727$ (Ensemble 2), $0.13754$ (Ensemble 3) for $u$, $d$-quarks while $\kappa_{s} = 0.13640$ (Each ensemble) for the $s$-quark.
For the charm quark, we employ the relativistic heavy quark (RHQ) action \cite{RHQ} to avoid the leading $\mathcal{O} \left( (m_Q a)^n \right)$ and the next-to-leading $\mathcal{O} \left( (m_Q a)^n (a \Lambda_{\mathrm{QCD}}) \right)$ discretization errors due to the charm quark mass $m_Q$. 
We use the RHQ parameters determined in Ref.~\cite{Namekawa} so as to reproduce the experimental value of the mass and the relativistic dispersion relation for the charmonium in spin-averaged $1S$ state. 
Note that charm-quark loops are not considered in the present paper. 
The effects of charm loops for charmed baryons are expected to be small as studied e.g. in Ref.~\cite{Charm_loop1, Charm_loop2}.
Nevertheless, the effects to the $\Lambda_c N$ interaction is an interesting open question to be investigated in the future.

We calculate quark propagators with the periodic boundary condition for the spatial directions, while the Dirichlet boundary condition is imposed on the temporal direction at the time-slice $t = 32 + t_0$.
Correlation functions for $\Lambda_c N$ are calculated using the unified contraction algorithm \cite{UCA_DOI}.
In order to increase the statistics, we take an average over forward and backward propagations in time. 
Furthermore, we take 64 different time-slices for each configuration as the wall source location. 
The total statistics of each ensemble are given in Table~\ref{table:statistics}. 
\begin{table}[h]
	\caption{The number of configurations, sources, and masses of pion and nucleon on each ensemble.
	The factor of two in \# of sources means forward and backward propagations in time. \label{table:statistics}}
	\begin{center}
		\begin{tabular}{c|cccc} \hline \hline
			& \# of gauge configs. & \# of sources & $m_\pi$ [MeV] & $m_N$ [MeV] \\ \hline
			Ensemble 1 & 399 & 64 $\times$ 2 & 702(2) & 1581(6) \\
			Ensemble 2 & 400 & 64 $\times$ 2 & 570(1) & 1399(9) \\
			Ensemble 3 & 450 & 64 $\times$ 2 & 412(2) & 1215(9) \\ \hline \hline
		\end{tabular}
	\end{center}
\end{table}

For all analyses in this study, we employ the jackknife method to estimate statistical errors. 
The bin-size of the jackknife samples is taken to 57, 40 and 45 for the Ensemble 1, 2 and 3, respectively. 
We confirm that change of bin-size does not affect the errors for hadron masses as well as the errors for potentials and phase shifts.

Various hadron masses calculated in this work are summarized in Table~\ref{table:baryon_masses}. 
\begin{table}[h]
	\caption{Calculated hadron masses in unit of [MeV] for each ensemble.
	The fit range in $t-t_0$ is [10, 20] for $\pi$ in Ensemble 2, [10, 15] for $\pi$ in Ensemble 3 and [15, 20] for all other cases. \label{table:baryon_masses}}
	\begin{center}
		\begin{tabular}{c|ccc} \hline \hline
			& Ensemble 1 & Ensemble 2 & Ensemble 3 \\ \hline
			$m_\pi$ & 702(2) & 570(1) & 412(2) \\
			$m_K$ & 789(2) & 713(1) & 637(2) \\
			$m_D$ & 1999(1) & 1949(2) & 1904(2) \\
			$m_N$ & 1581(6) & 1399(9) & 1215(9) \\
			$m_\Lambda$ & 1642(6) & 1493(7) & 1342(6) \\
			$m_\Sigma$ & 1657(6) & 1522(8) & 1395(9) \\
			$m_{\Sigma^*}$ & 1881(11) & 1749(16) & 1631(28) \\
			$m_{\Lambda_c}$ & 2685(3) & 2555(5) & 2434(6) \\
			$m_{\Sigma_c}$ & 2780(5) & 2674(7) & 2575(9) \\
			$m_{\Sigma^*_c}$ & 2866(5) & 2763(7) & 2661(10) \\ \hline \hline
		\end{tabular}
	\end{center}
\end{table}

\section{Numerical results} \label{sec:results_pot}
\subsection{$\Lambda_c N$ central potentials in $^1S_0$ channel}
We first discuss the central potential for the $S$-wave spin-singlet $\Lambda_c N$ system.
In order to obtain the potential, we use the $R$-correlator with the $J^P=0^+$ wall source defined in Eq.~(\ref{eq:Rcorr}), which is further projected to the $^1S_0$ channel as
\begin{equation}
	R_{^1S_0} (\vec{r}, t-t_0) \equiv P^{(L=0)} P^{(S=0)}_{\beta \alpha} R_{\alpha \beta} (\vec{r}, t-t_0; J^P = 0^+),
	\label{eq:Rcorr_1S0}
\end{equation}
where $P^{(S=0)}_{\beta \alpha}$ and $P^{(L=0)}$ are projection operators to the total spin $S=0$ and the orbital angular momentum $L=0$, respectively\footnote{These projections are redundant since the $J^P = 0^+$ state allows only the $^1S_0$ channel.}. 
On the lattice, we employ the cubic transformation group for the projection of the orbital angular momentum as 
\begin{equation}
	P^{(L=0)} R (\vec{r}, t-t_0) \equiv \frac{1}{24} \sum_{g \in SO(3, \mathbb{Z})} R (g^{-1} \vec{r}, t-t_0), \label{eq:Proj_L}
\end{equation}
where $g$ is one of 24 elements in $SO(3, \mathbb{Z})$. 
This projection picks up an $A^+_1$ representation of $SO(3, \mathbb{Z})$.
By using the $R$-correlator in the $^1S_0$ channel, we extract the $\Lambda_c N$ central potential through Eq.~(\ref{eq:time-dep_HAL}).
Since $\vec{\sigma_1} \cdot \vec{\sigma_2} = -3$ and $S_{12} = 0$ for the  $J^P = 0^+$ state, we have 
\begin{eqnarray}
	V^{(0^+)}_C (\vec{r}) &\equiv& V_0 (\vec{r}) - 3 V_\sigma (\vec{r}) \nonumber \\
	&=& \frac{1}{R_{^1S_0} (\vec{r}, t-t_0)} \left[ \left( \frac{1 + 3 \delta^2}{8 \mu} \right) 
	\frac{\partial^2}{\partial t^2} -\frac{\partial}{\partial t} - H_0 \right] R_{^1S_0} (\vec{r}, t-t_0)\label{eq:pot_1S0}
\end{eqnarray}
for the moderately large $t-t_0$.

\begin{figure}[h] \centering
	\includegraphics[width=13cm]{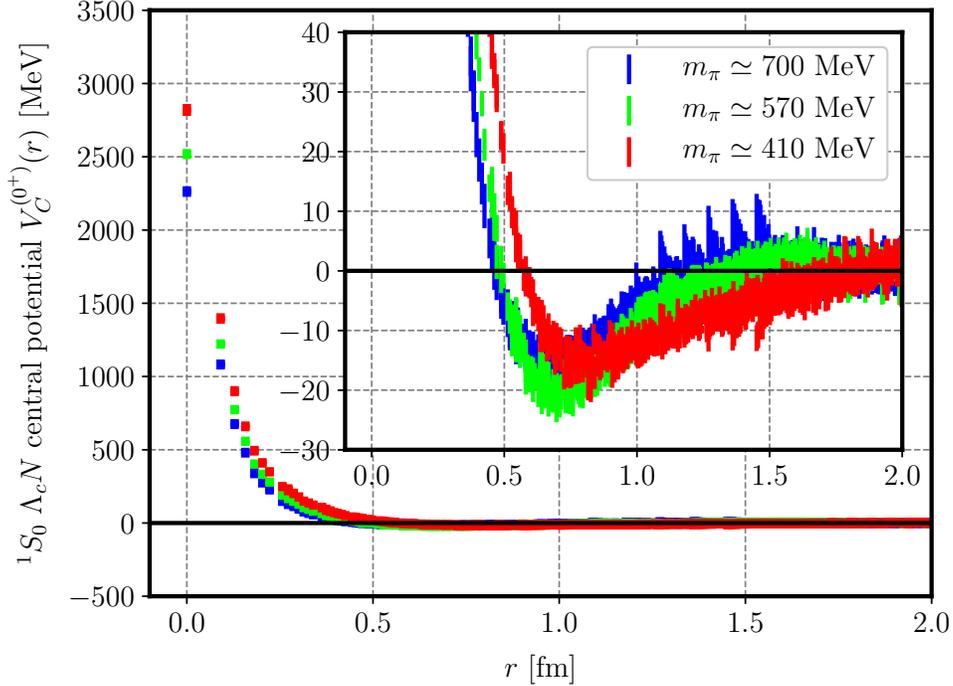} 
	\caption{The $\Lambda_c N$ central potential in the $^1S_0$ channel for each ensemble. 
	The potential is calculated at $t-t_0 = 13$ for $m_{\pi}\simeq $~700 MeV case (Blue), $t-t_0 = 11$ for $m_{\pi}\simeq $~570 MeV case (Green) and $t-t_0 = 9$ for $m_{\pi}\simeq $~410 MeV case (Red). \label{fig:LcN_potential_1S0}}
\end{figure}
Fig.~\ref{fig:LcN_potential_1S0} shows the $\Lambda_c N$ central potential in the $^1S_0$ channel for each ensemble. 
The potential is calculated at $t-t_0=13$ (Ensemble 1: $m_{\pi}\simeq $~700 MeV), $t-t_0=11$ (Ensemble 2: $m_{\pi}\simeq $~570 MeV) and $t-t_0=9$ (Ensemble 3: $m_{\pi}\simeq $~410 MeV). 
We find a repulsive core at short distances ($r \lesssim 0.5$ fm) and an attractive pocket at intermediate distances ($0.5 \lesssim r \lesssim 1.5$ fm) in the $\Lambda_c N$ potential. 
We also observe that the height of the repulsive core increases and the minimum of the attractive pocket shifts outward, as $u$, $d$ quark masses decrease. 
A variation of the repulsive core against $u$, $d$ quark masses may be explained by the fact that the color magnetic interaction is proportional to the inverse of the constituent quark mass \cite{CMI}. 
We notice that the attraction of the $\Lambda_c N$ potential seems weaker than that of the $\Lambda N$ potential in Ref.~\cite{Nemura_LN}.

\begin{figure}[h] \centering
	\includegraphics[width=17cm]{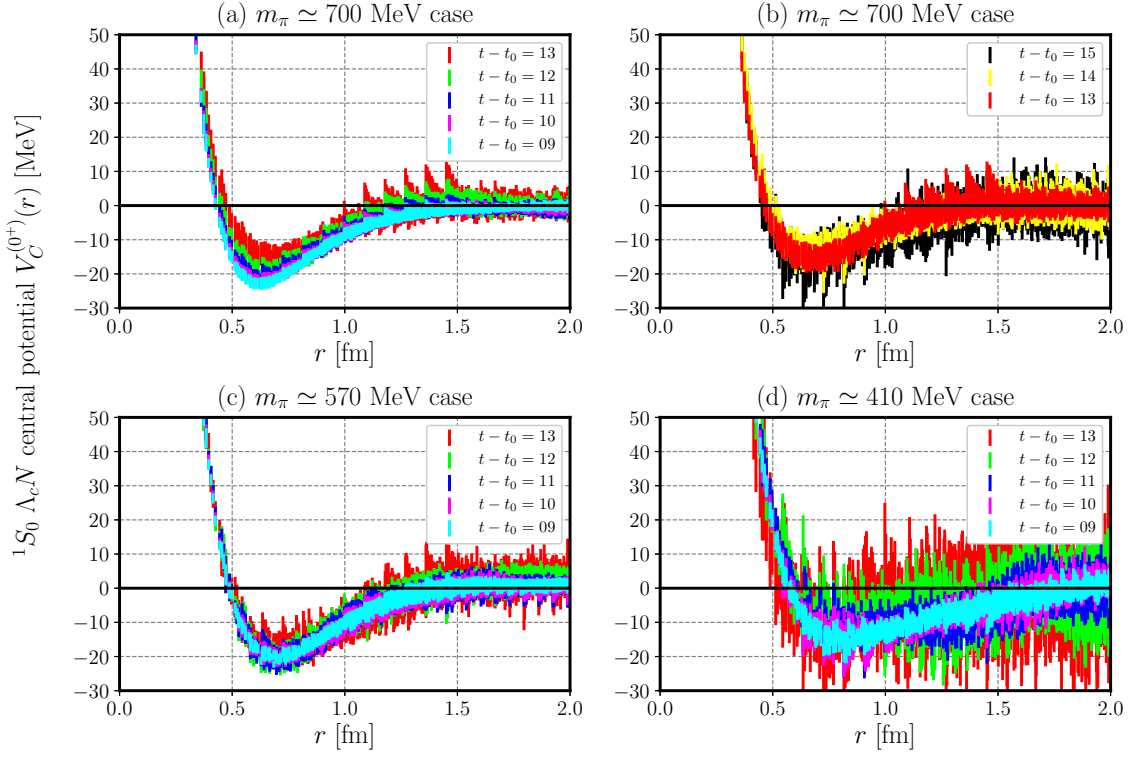} 
	\caption{Time dependence of the $\Lambda_c N$ central potential in the $^1S_0$ channel for $m_{\pi}\simeq $~700 MeV case (a, b), 
	$m_{\pi}\simeq $~570 MeV case (c) and $m_{\pi}\simeq $~410 MeV case (d). \label{fig:LcN_potential_Tdep}}
\end{figure}
In order to check the stability of the potential against the time separations $t-t_0$, we plot the time dependence of the $\Lambda_c N$ central potential in the $^1S_0$ channel at several different time separation $t-t_0$ in Fig.~\ref{fig:LcN_potential_Tdep}. 
In Fig.~\ref{fig:LcN_potential_Tdep} (a), we find that the potential exhibits non-negligible time dependence for $t-t_0 \in [9, 12]$. 
The potential, however, becomes stable at larger time, $t-t_0 \in [13, 15]$, as shown in Fig.~\ref{fig:LcN_potential_Tdep} (b).
This observation suggests that contributions from inelastic states are negligible at $t-t_0 \ge 13$, so that we take the potential at $t-t_0 = 13$ for $m_{\pi}\simeq $~700 MeV case.
For $m_{\pi}\simeq $~570 and 410 MeV cases, we find that potentials are stable within the statistical errors at earlier time slices ($t-t_0 \in [9, 13]$) as shown in Fig.~\ref{fig:LcN_potential_Tdep} (c,d).
This implies that the contaminations from inelastic states are more suppressed as $ud$ quark masses are decreased: 
In fact, the excitation energy to the lowest inelastic state ($\Sigma_c N$) becomes larger for lighter $ud$ quark masses.
Another possible reason is that the statistical errors at fixed $t-t_0$ become larger for lighter quark masses and tend to dominate the total error budget compared to the systematic errors from inelastic states.
Under these considerations, we take $t-t_0 = 11$ at $m_{\pi}\simeq $~570 MeV and $t-t_0 = 9$ at $m_{\pi}\simeq $~410 MeV in the following analyses. 

We then calculate physical observables such as scattering phase shifts in the $^1S_0$ channel from the potential. 
For this purpose, we fit the potential data with the functional form given by
\begin{eqnarray}
	V_{\mathrm{fit}} (r) = a_1 e^{-\left( \frac{r}{a_2} \right)^2} + a_3 e^{-\left( \frac{r}{a_4} \right)^2} 
	+ a_5 \left[ \left( 1 - e^{-a_6 r^2} \right) \frac{e^{-a_7 r}}{r} \right]^2. \label{eq:fit_function_2G1Ysq}
\end{eqnarray}
Table~\ref{table:fit_params} shows fit-parameters for the $\Lambda_c N$ central potential in the $^1S_0$ channel on each ensemble. 
Using data at $r \in (0.0, 2.0]$ fm, we achieve $\chi^2/\mathrm{dof} \simeq 1$.
\begin{table}[h]
	\caption{Fit parameters of $V_{\mathrm{fit}} (r)$ defined in Eq.~(\ref{eq:fit_function_2G1Ysq}) for the $\Lambda_c N$ (effective) central potential, where $a_1$ and $a_3$ are expressed in unit of [MeV], $a_2$ and $a_4$ are expressed in unit of [fm], $a_5$, $a_6$ and $a_7$ are expressed in unit of [MeV~fm$^2$], [fm$^{-2}$] and [fm$^{-1}$], respectively. \label{table:fit_params}}
	\begin{center}
		\scalebox{0.87}{
		\begin{tabular}{c|rrr|rrr} \hline \hline
			& & $^1S_0$ channel & & & $^3S_1$ channel & \\ \hline
			$m_{\pi}$ & 702(2) MeV & 570(1) MeV & 412(2) MeV & 702(2) MeV & 570(1) MeV & 412(2) MeV \\ \hline
			$a_1$ & 1090(36) & 1266(20) & 1520(24) & 458.1(53.8) & 682.6(13.4) & 853.8(17.2) \\
			$a_2$ & 0.09761(233) & 0.09912(112) & 0.1121(11) & 0.09296(835) & 0.1061(138) & 0.1183(16) \\
			$a_3$ & 854.4(50.2) & 892.5(27.3) & 712.4(12.8) & 761.6(71.8) & 631.0(17.5) & 569.2(11.2) \\
			$a_4$ & 0.4384(45) & 0.4670(36) & 0.6808(51) & 0.4208(59) & 0.4886(32) & 0.6898(56) \\
			$a_5$ & -18637(5796) & -29804(6231) & -45479(4116) & -71142(38550) & -19158(3687) & -40798(3994) \\
			$a_6$ & 1.566(154) & 1.182(84) & 0.6635(229) & 0.8462(1626) & 1.163(77) & 0.6144(221) \\
			$a_7$ & 3.493(122) & 3.308(74) & 2.367(25) & 3.971(164) & 3.071(66) & 2.331(27) \\ \hline \hline
		\end{tabular}
		}
	\end{center}
\end{table}
We solve the Schr\"odinger equation with the fitted potential in the infinite volume and extract its phase shifts from the asymptotic behavior of the wave function. Finally, the $S$-wave scattering length is calculated as
\begin{eqnarray}
	a = \lim_{k \to 0} \frac{\tan \delta_{00} (k)}{k}. \label{eq:scattering_length}
\end{eqnarray}
Here we employ the particle physics convention for the definition of scattering length which has opposite sign from the historical sign convention of the baryon-baryon interaction.

\begin{figure}[h] \centering
	\includegraphics[width=16cm]{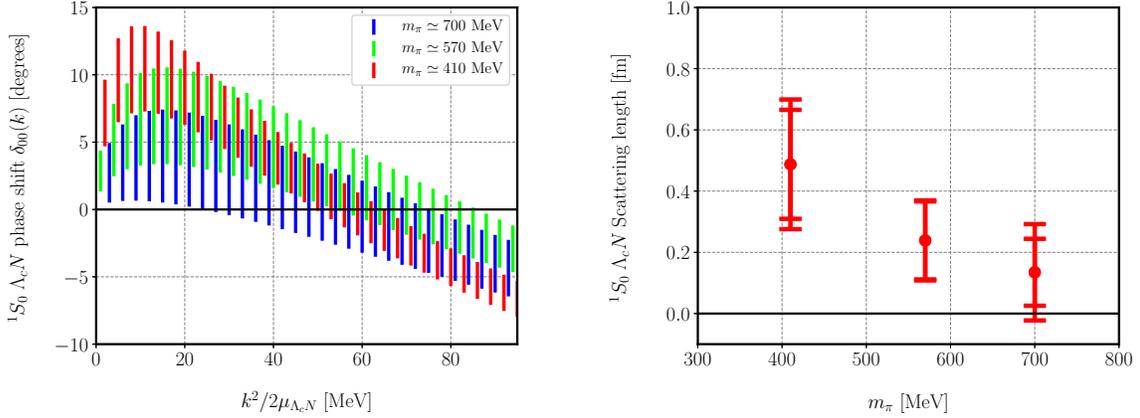} 
	\caption{The phase shift (left) and the scattering length (right) for the $\Lambda_c N$ system in the $^1S_0$ channel with particle physics sign convention of the scattering length.
	The inner error of the scattering length is statistical only, while the outer represents the total one (statistical and systematic errors added in quadrature). \label{fig:LcN_phase_1S0}}
\end{figure}
\begin{table}[h] 
	\caption{The scattering length for the $\Lambda_c N$ system with the particle physics sign convention.
	The first parenthesis indicates the statistical error, and the second parentheses indicates the systematic error evaluated by the difference between the mean value at $t-t_0$ and that at $t-t_0+2$. 
	\label{table:LcN_scatt}}
	\begin{center}
		\begin{tabular}{c|c|c} \hline \hline
			$m_\pi$ & $^1S_0$ channel &$^3S_1$ channel  \\ \hline
			412(2) MeV & 0.49 (18) (11) fm & 0.51 (20) (~9) fm \\
			570(1) MeV & 0.24 (13) (~3) fm & 0.29 (16) (~9) fm \\
			702(2) MeV & 0.13 (11) (11) fm & 0.17 (10) (12) fm \\ \hline \hline
		\end{tabular}
	\end{center}
\end{table}
Fig.~\ref{fig:LcN_phase_1S0} show the phase shift and the scattering length for the $\Lambda_c N$ system in the $^1S_0$ channel for each ensemble, and the numerical values of the scattering length are listed in Table~\ref{table:LcN_scatt}.
Systematic errors of the scattering length are evaluated by the difference between the mean value at $t-t_0$ and that at $t-t_0+2$, where $t-t_0=13$, $11$ and $9$ for $m_{\pi}\simeq $~700, 570 and 410 MeV case, respectively. 
Results of these observables indicate that the net interaction in the $^1S_0$ channel is attractive at low energies ($E \lesssim 40$ MeV) in all cases, but not strong enough to form bound states.
We also notice a tendency that the attraction becomes stronger as the pion mass decreases
\footnote{We have reported preliminary results of the $\Lambda_c N$ potential at $t-t_0=9$ in Ref.~\cite{PoS2015_LcN} and those at $t-t_0=10$ in Ref.~\cite{PoS2016_LcN}. 
Since we have more statistics than those preliminary studies, we could analyze the potential at larger $t-t_0$ in the present paper. 
As a consequence, together with the large statistical fluctuation observed at $m_\pi \simeq 410$ MeV in the $^3S_1$ channel, the preliminary results show an opposite tendency in terms of the quark mass dependence of the scattering lengths.}.

The leading order approximation of the $\Lambda_cN$ potential may have sizable systematic errors once $E$ approaches to the $\Sigma_c N$ threshold from below ($E \simeq 96, 121 $ and $145$ MeV for $m_\pi \simeq 700, 570$ and $410$ MeV case, respectively) due to the truncation of the derivative expansion of the non-local potential.
Such systematic uncertainties of the $\Lambda_cN$ interaction near the $\Sigma_cN$ threshold 
can be estimated by comparing the phase shift calculated by the leading order potential and that
 obtained by the the coupled-channel potential \cite{HAL_CCP, Sasaki}.
Our preliminary results of the coupled-channel potential in Ref.~\cite{PoS2016_LcN} indicate
 that the phase shifts obtained by the two methods in the $^3S_1$ channel for $\Lambda_c N$ system
agree with each other even near the $\Sigma_c N$ threshold. More systematic study with 
both $^3S_1$ and $^1S_0$ channels  are needed to draw quantitative conclusion.

\subsection{$\Lambda_c N$ central and tensor potentials in $^3S_1-^3D_1$ channel}
The $J^P = 1^+$  wall source in Eq.~(\ref{eq:wall_source}) generates states in the $^3S_1$ channel at $t_0$, but the $R$-correlator contains both $^3S_1$ and $^3D_1$ channels at $t > t_0$ due to QCD interactions.
This fact is translated into the existence of the tensor potential, which is the transition potential between these two channels. 
To extract both central and tensor potentials, we introduce the projections to the $S$-wave and $D$-wave components as
\begin{eqnarray}
	P_S R_{\alpha \beta} (\vec{r}, t-t_0) &\equiv& P^{(L=0)} R_{\alpha \beta} (\vec{r}, t-t_0) \nonumber \\
	P_D R_{\alpha \beta} (\vec{r}, t-t_0) &\equiv& \left( 1 - P^{(L=0)} \right) R_{\alpha \beta} (\vec{r}, t-t_0), 
	\label{eq:Rcorr_SDproj}
\end{eqnarray}
where $P^{(L=0)}$ is the projection operator defined in Eq.~(\ref{eq:Proj_L}). 
Using Eq.~(\ref{eq:Rcorr_SDproj}), both central and tensor potentials are extracted from the coupled channel Schr\"odinger equation as
\begin{eqnarray}
	\mathcal{K} \left[ P_S R_{\alpha \beta} (\vec{r}, t-t_0) \right] 
	&=& V_C^{(1^+)} (\vec{r}) \left[ P_S R_{\alpha \beta} (\vec{r}, t-t_0) \right]
	+ V_T(\vec{r}) \left[ P_S \left( S_{12} R \right)_{\alpha \beta} (\vec{r}, t-t_0) \right], \nonumber \\
	\mathcal{K} \left[ P_D R_{\alpha \beta} (\vec{r}, t-t_0) \right] 
	&=& V_C^{(1^+)} (\vec{r}) \left[ P_D R_{\alpha \beta} (\vec{r}, t-t_0) \right]
	+ V_T(\vec{r}) \left[ P_D \left( S_{12} R \right)_{\alpha \beta} (\vec{r}, t-t_0) \right], \nonumber \\
	\label{eq:Sch_tensor}
\end{eqnarray}
where $\mathcal{K}$ is the operator in the left-hands of Eq.~(\ref{eq:time-dep_HAL}), given by
\begin{eqnarray}
	\mathcal{K} \equiv \left( \frac{1 + 3 \delta^2}{8 \mu} \right) \frac{\partial^2}{\partial t^2} -\frac{\partial}{\partial t} - H_0,
\end{eqnarray}
and the central potential with $J^P=1^+$ is expressed as
\begin{eqnarray}
	V_C^{(1^+)} (\vec{r}) = V_0 (\vec{r}) + V_\sigma (\vec{r}), \label{eq:pot_3S1}
\end{eqnarray}
since $(\vec{\sigma_1} \cdot \vec{\sigma_2}) = +1$. 

\begin{figure}[h] \centering
	\includegraphics[width=16cm]{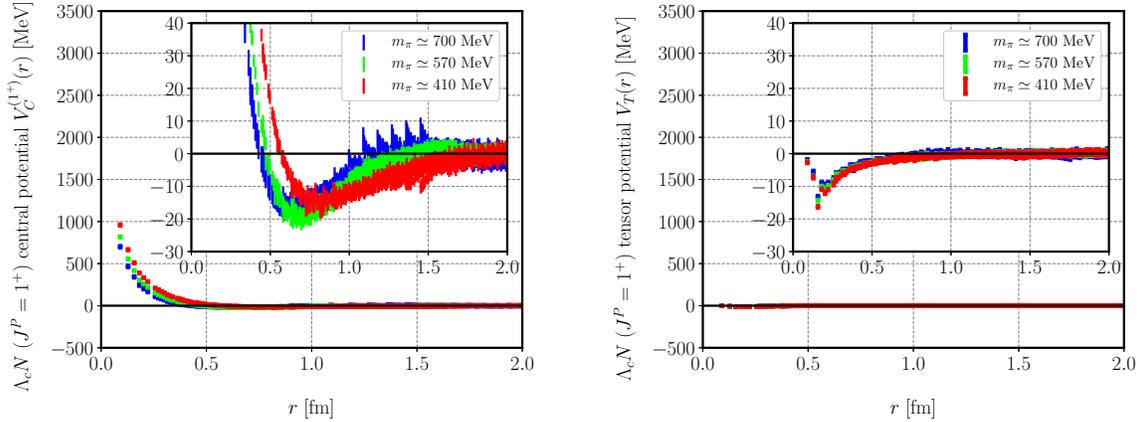} 
	\caption{$\Lambda_c N$ central potential (left) and tensor potential (right) in the $J^P=1^+$ state for each ensemble. 
	The potentials are calculated at $t-t_0 = 13$ for $m_{\pi}\simeq $~700 MeV case (Blue), $t-t_0 = 11$ for $m_{\pi}\simeq $~570 MeV case (Green) and $t-t_0 = 9$ for $m_{\pi}\simeq $~410 MeV case (Red). \label{fig:LcN_potential_tensor}}
\end{figure}
Fig.~\ref{fig:LcN_potential_tensor} shows that the central potential (left) and the tensor potential (right) for the $\Lambda_c N$ system with $J^P=1^+$. 
These potentials are calculated at $t-t_0 = 13$ (Ensemble 1: $m_{\pi}\simeq$ 700 MeV), $t-t_0 = 11$ (Ensemble 2: $m_{\pi}\simeq$ 570 MeV) and $t-t_0 = 9$ (Ensemble 3: $m_{\pi}\simeq$ 410 MeV). 
It is confirmed that these potentials are stable against the change of $t-t_0$ within the statistical errors, as was observed in the central potential in the $^1S_0$ channel. 
We notice that the central potential is similar to the one in the $^1S_0$ channel except at short distances ($r \lesssim 0.5$ fm). 
The tensor potential of the $\Lambda_c N$ system is weak compared to that of the $\Lambda N$ system \cite{Nemura_LN}. 
We also find that the $u$, $d$ quark mass dependence of the tensor potentials is weak. 

The weaker $\Lambda_c N$ potential than $\Lambda N$ could be explained from following facts:
(i) The long-range contribution is expected to be caused by the $K$ meson exchange for $\Lambda N$ interaction \cite{Bando}. 
In the $\Lambda_c N$ system, however, the $K$ meson (strange quark) exchange is replaced by the $D$ meson (charm quark) exchange, and this contribution is highly suppressed due to the much heavier $D$ meson mass than the $K$ meson mass.
(ii) The one-pion exchange in the $\Lambda N - \Sigma N$ transition is considered to give a sizable contribution to the effective $\Lambda N$ interaction. 
In the $\Lambda_c N$ system, however, this contribution is expected to be suppressed due to the large mass difference between $\Lambda_c N$ and $\Sigma_c N$.

By the same procedure as in the case of the $^1S_0$, we calculate the phase shifts and the scattering lengths in this system. 
Since the tensor potential is shown to be weak, we employ the effective central potential in the $^3S_1$ channel (instead of the  $^3S_1$ - $^3D_1$ coupled channel), which implicitly includes the effect of the tensor potential through virtual processes such as $^3S_1\to$ $^3D_1\to$ $^3S_1$.
\begin{figure}[h] \centering
	\includegraphics[width=13cm]{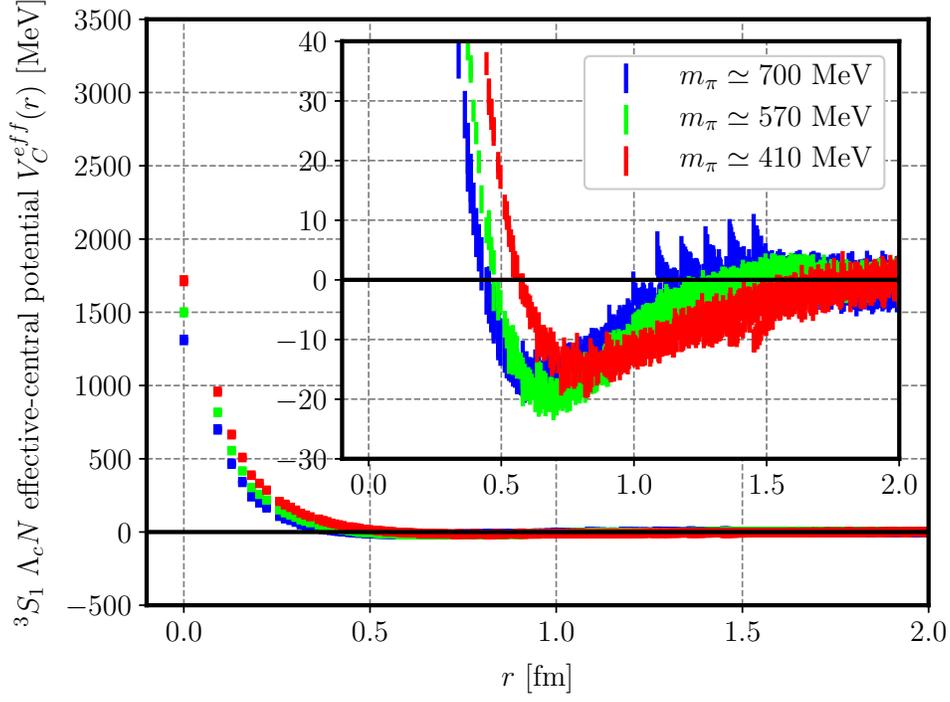}
	\caption{The $\Lambda_c N$ effective-central potential in the $^3S_1$ channel for each ensemble. 
	The potential is calculated at $t-t_0 = 13$ for $m_{\pi}\simeq$ 700 MeV case (Blue), $t-t_0 = 11$ for $m_{\pi}\simeq$ 570 MeV case (Green) and $t-t_0 = 9$ for $m_{\pi}\simeq$ 410 MeV case (Red). \label{fig:LcN_potential_3S1}}
\end{figure}
\begin{figure}[h] \centering
	\includegraphics[width=16cm]{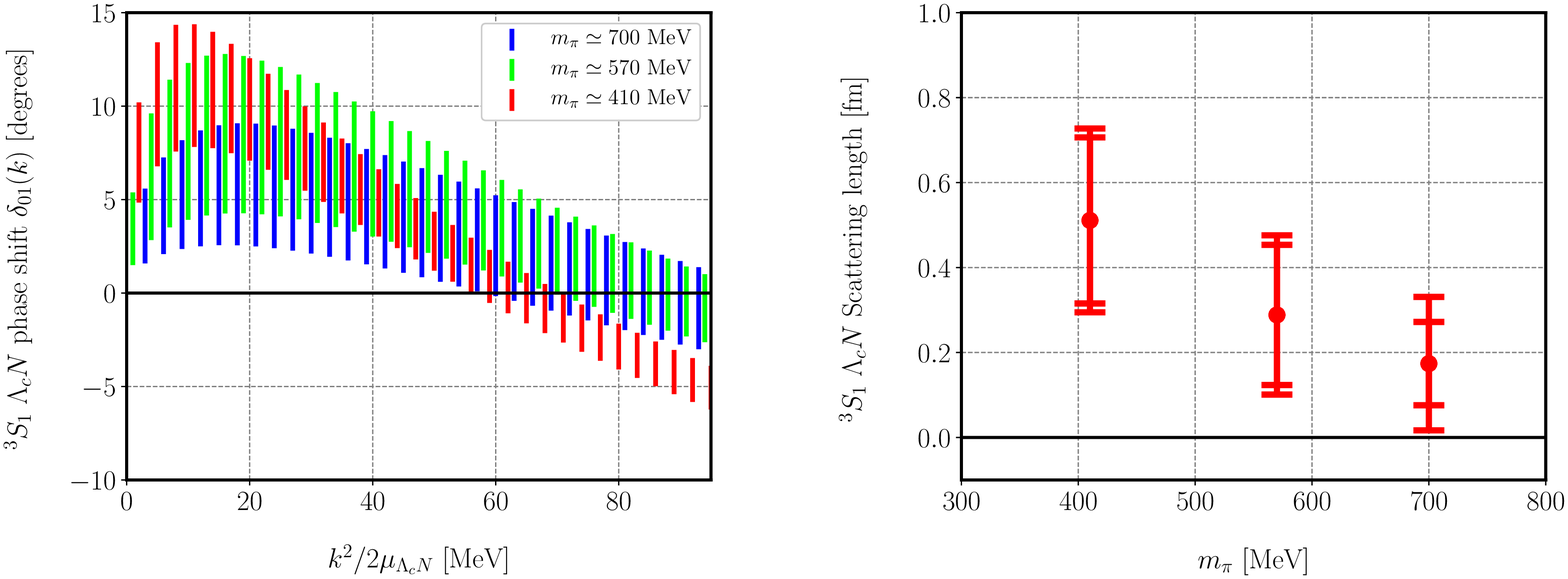} 
	\caption{The phase shifts (left) and the scattering length (right) for the $\Lambda_c N$ system in the $^3S_1$ channel with particle physics sign convention of the scattering length.
	The inner error of the scattering length is statistical only, while the outer represents the total one (statistical and systematic errors added in quadrature). \label{fig:LcN_phase_3S1}}
\end{figure}
The $\Lambda_c N$ effective-central potential is plotted in Fig.~\ref{fig:LcN_potential_3S1}, and fit parameters are given in Table~\ref{table:fit_params}. 
The phase shift and the scattering length shown in Fig.~\ref{fig:LcN_phase_3S1} are very similar to those in the $^1S_0$ channel.
See also Table~\ref{table:LcN_scatt} for a comparison of the scattering length between two channels. 

\subsection{Spin independence of central potentials}
\begin{figure}[h] \centering
	\includegraphics[width=16cm]{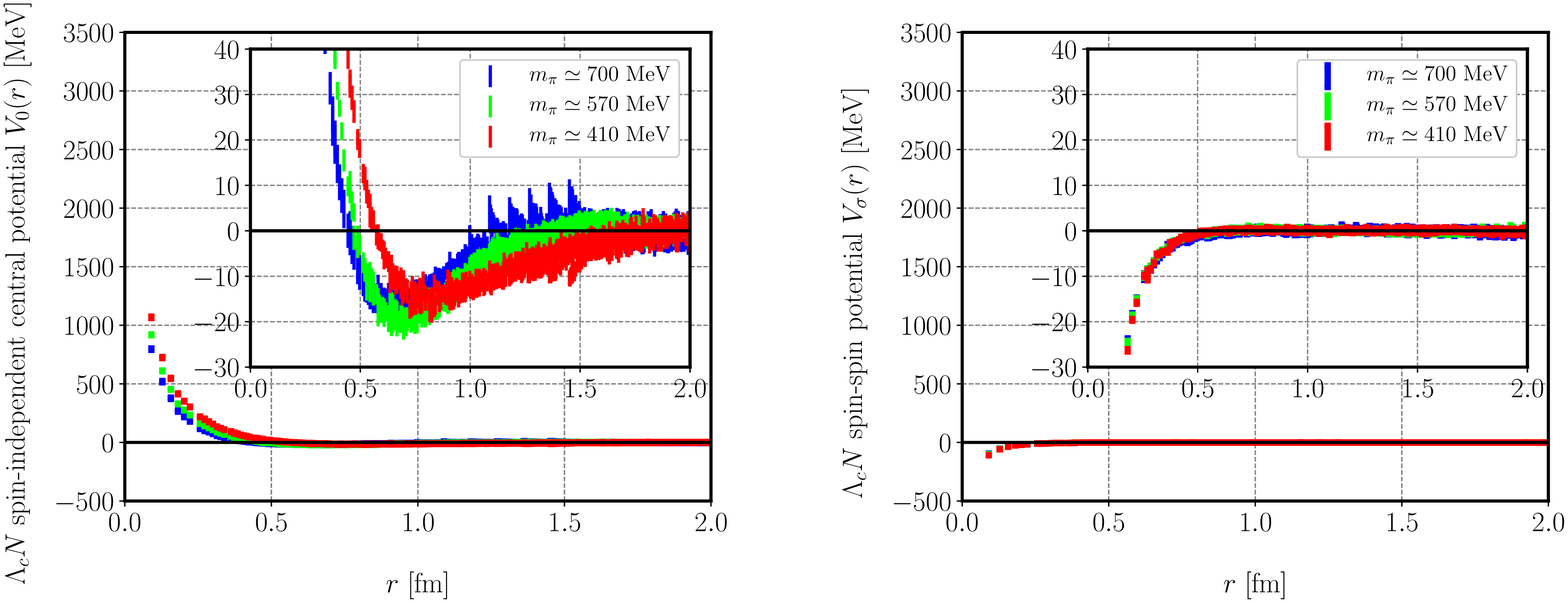}
	\caption{$\Lambda_c N$ spin-independent central potential $V_0$ (left) and the spin-dependent one $V_\sigma$ (right) for each ensemble. 
	The potentials are calculated at $t-t_0 = 13$ for $m_{\pi}\simeq$ 700 MeV case (Blue), $t-t_0 = 11$ for $m_{\pi}\simeq$ 570 MeV case (Green) and $t-t_0 = 9$ for $m_{\pi}\simeq$ 410 MeV case (Red). \label{fig:LcN_potential_spin_spin}}
\end{figure}
In this subsection, we quantify a similarity between the $^1S_0$ central potential and the $^3S_1$ effective central potential. 
For this purpose, we further decompose the central potential in both $J^P = 0^+$ state (Eq.~\ref{eq:pot_1S0}) and $J^P = 1^+$ state (Eq.~\ref{eq:pot_3S1}) into the spin-independent central potential $V_0$ and the spin-dependent one $V_\sigma$, which are extracted as
\begin{eqnarray}
	V_0 (\vec{r}) = \frac{1}{4} \left( 3 V_C^{(1^+)} (\vec{r}) + V_C^{(0^+)} (\vec{r}) \right) \\
	V_\sigma (\vec{r}) = \frac{1}{4} \left( V_C^{(1^+)} (\vec{r}) - V_C^{(0^+)} (\vec{r}) \right).
\end{eqnarray}
Fig.~\ref{fig:LcN_potential_spin_spin} shows the $\Lambda_c N$ spin-independent central potential $V_0$ (left) and the spin-dependent one $V_\sigma$ (right).
It is easy to see that the spin-dependent potential $V_\sigma$ is negligibly small, and the spin-independent central potential gives a significant contribution for $\Lambda_c N$ potentials.
The origin of the small spin-dependent potential $V_\sigma$ could be explained by the heavy D-meson mass and the large separation between $\Lambda_cN$ and $\Sigma_cN$, similar to the case of $\Lambda_cN$ tensor potential in $^3S_1$-$^3D_1$ channel.

\section{Possible $\Lambda_c$ hypernuclei} \label{sec:folding_pot}
Since the $\Lambda_c N$ interaction is dominated by the spin-independent central force, as we discussed in the previous section, the spectrum of $\Lambda_c$ hypernuclei, if they exist, would be simple.
In order to investigate $\Lambda_c$ hypernuclei, we employ the single-folding potential defined by
\begin{eqnarray}
	V_F (\vec{r}) = \int d^3 r^\prime \rho_A (\vec{r^\prime}) V_{\Lambda_c N} (\vec{r} - \vec{r^\prime}), \label{eq:Fpot}
\end{eqnarray}
where $\rho_A (\vec{r})$ denotes nuclear density distributions with the atomic number $A$, and $V_{\Lambda_c N} (\vec{r})=V_0(\vec r)$ stands for the two body spin-independent central potential of the $\Lambda_c N$ system. 

For the nuclear density distribution function, we use the two-parameter Fermi form given by
\begin{eqnarray}
	\rho_A (\vec{r}) = \rho_0 \left[ 1 + \exp\left( \frac{r - c}{a} \right) \right]^{-1}, \hspace{1cm} \int d^3 r~\rho_A (\vec{r}) = A,
\end{eqnarray}
where $r \equiv |\vec{r}|$. We employ the parameters $\rho_0$, $c$, $a$ given in Ref.~\cite{SFP_params} for spherical nuclei such as $^{12}$C, $^{28}$Si, $^{40}$Ca, $^{58}$Ni, $^{90}$Zr and $^{208}$Pb.

\begin{figure}[h] \centering
	\includegraphics[width=13cm]{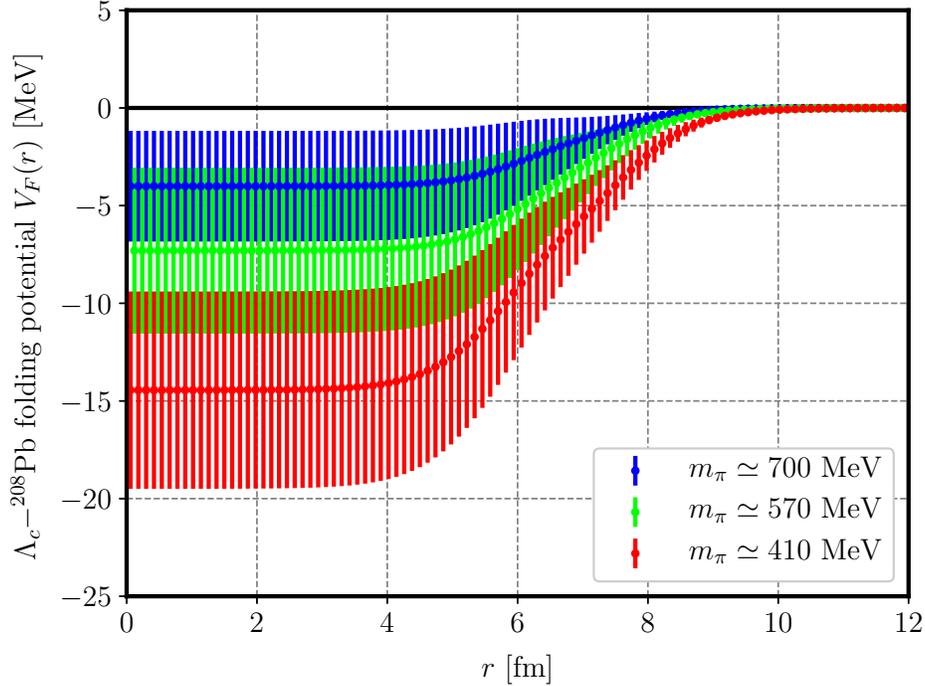} 
	\caption{$\Lambda_c - ^{208}$Pb folding potentials calculated from the spin-independent central potential of the $\Lambda_c N$ system (Fig.~\ref{fig:LcN_potential_spin_spin}) for $m_{\pi}\simeq $~700, 570 and 410 MeV cases. \label{fig:LcN_Fpot_Pb}}
\end{figure}
Fig.~\ref{fig:LcN_Fpot_Pb} shows the folding potential for $\Lambda_c-^{208}$Pb for each ensemble.
We observe that the folding potential becomes deeper as the $u$, $d$ quark masses decrease and becomes as large as $-10$ to $-20$ MeV at the origin.

Using this folding potential, we calculate the binding energy of the $\Lambda_c$ hypernuclei by the Gaussian expansion method \cite{Hiyama_GEM} for the $S$-wave potential, with the physical masses for $\Lambda_c$ and nuclei.
\begin{figure}[h] \centering
	\includegraphics[width=13cm]{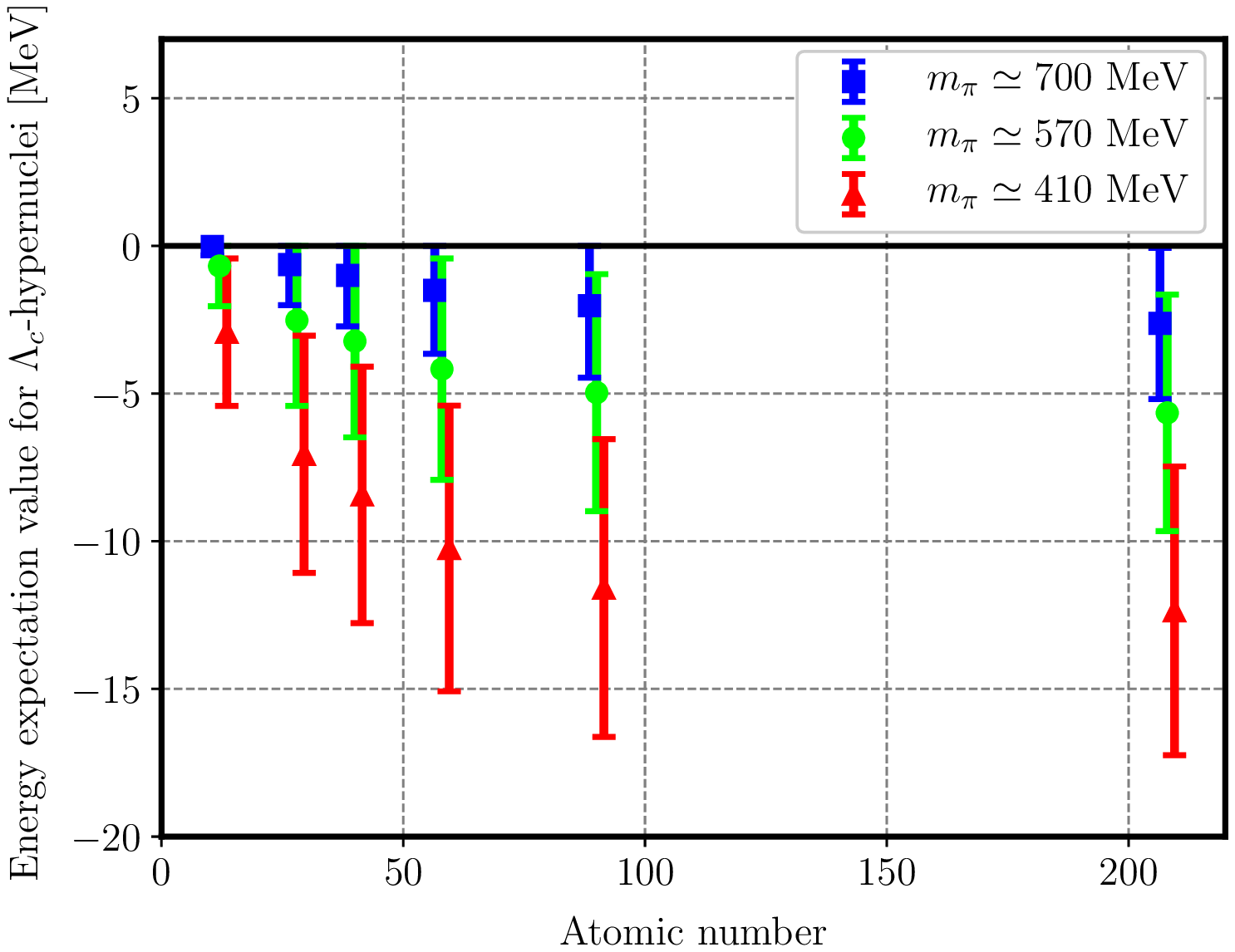} 
	\caption{The binding energy of $\Lambda_c$ in symmetric nuclei such as $^{12}$C, $^{28}$Si, $^{40}$Ca, $^{58}$Ni, $^{90}$Zr and $^{208}$Pb for each ensemble. 
	The binding energies are calculated from the folding potentials for $\Lambda_c$ hypernuclei by using the Gaussian expansion method.
	The folding potentials are constructed from the spin-independent central potential of the $\Lambda_c N$ system (Fig.~\ref{fig:LcN_potential_spin_spin}) for $m_{\pi}\simeq $~700, 570 and 410 MeV cases. 
	In the calculation of the binding energies, we adjust the mass of $\Lambda_c$ and nuclei to those of physical values. \label{fig:LcN_Fpot_Eb}}
\end{figure} 
Fig.~\ref{fig:LcN_Fpot_Eb} shows the binding energy of several $\Lambda_c$ hypernuclei for each ensemble. 
As we expected, the binding energy $|E_b|$ increases as the atomic number increases. 
Furthermore, as the $\Lambda_c N$ potential approaches to the physical one (as the $u$, $d$ quark masses decrease toward physical values), the binding energy increases.
These results suggest that $\Lambda_c$ hypernuclei may exist, if their binding energy is larger than the Coulomb repulsion. 
In order to estimate the effect of Coulomb force, we calculate the expectation value for the Coulomb potential using the binding solutions of $\Lambda_c$ hypernulei $| \psi_b \rangle$ as
\begin{eqnarray}
	E_{\mathrm{Coulomb}} = \frac{\langle \psi_b | V^C_F | \psi_b \rangle}{\langle \psi_b | \psi_b \rangle}, 
\end{eqnarray}
where $V^C_F$ is the single-folding Coulomb potential defined by 
\begin{eqnarray}
	V^C_F (\vec{r}) &=& \int d^3 r^\prime \rho_{\mathrm{ch}} (\vec{r^\prime}) V_{\mathrm{Coulomb}} (\vec{r} - \vec{r^\prime}),
\end{eqnarray}
where $V_{\mathrm{Coulomb}}(\vec{r})$ is an ordinary Coulomb potential and $\rho_{\mathrm{ch}}$ is charge density distribution by the Fourier-Bessel coefficient obtained from elastic electron scattering \cite{Charge_dens}.
\begin{figure}[h] \centering
	\includegraphics[width=13cm]{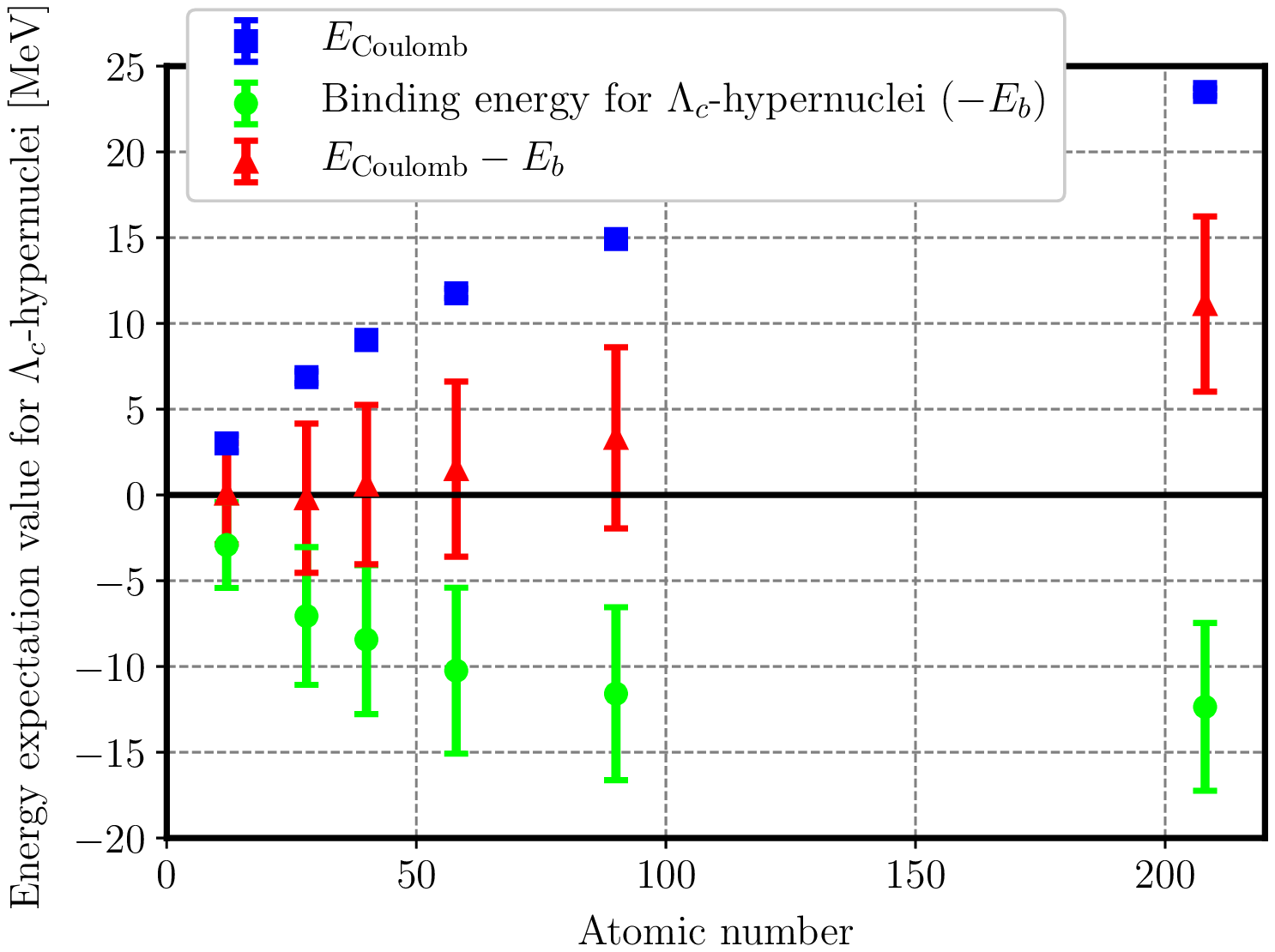} 
	\caption{The expectation value of folding potential for Coulomb force in $\Lambda_c$ hypernuclei (Blue). 
	The expectation values are calculated from the binding solution of the $\Lambda_c$ hypernuclei for Ensemble 3 ($m_{\pi}\simeq$ 410 MeV).
	For comparison, the binding energy of $\Lambda_c$ hypernuclei (Green) and sum of them (Red) are also plotted. \label{fig:LcN_Fpot_Eb+Ec}}
\end{figure} 
Fig.~\ref{fig:LcN_Fpot_Eb+Ec} shows the expectation values of the folding potential for Coulomb force calculated by using the binding solution of $\Lambda_c$ hypernuclei for Ensemble 3 ($m_{\pi}\simeq$ 410 MeV). 
For comparison, we also plot the binding energy for $\Lambda_c$ hypernuclei without Coulomb potential and the sum of them in Fig.~\ref{fig:LcN_Fpot_Eb+Ec}.
We observe that the Coulomb repulsion is large for heavy nuclei and $\Lambda_c-^{208}$Pb state becomes unbound with Coulomb force.
In the nuclei for $A=12-58$, on the other hand, the expectation values of Coulomb force are not much stronger than the binding energy of $\Lambda_c$ hypernuclei.
Since the binding energy increases as the attraction of the $\Lambda_c N$ potential becomes stronger toward the physical quark mass, this observation suggests a possibility that $\Lambda_c$ hypernuclei may exist in light or medium-heavy nuclei.

\section{Summary and conclusions} \label{sec:summary}
We have investigated the $\Lambda_c N$ interaction on the basis of lattice QCD simulations.  
The potentials have been extracted by the HAL QCD method using the (2+1)-flavor full QCD configurations with the lattice volume of $(2.9~\mathrm{fm})^3$ and the pion mass, $m_\pi \simeq 410,~570,~700$ MeV. 
We have extracted the central potential in $^1S_0$ channel and the central and tensor potential in $^3S_1-$$^3D_1$ channel. 
We found a repulsion at short distances and an attraction at intermediate distances in the central potentials for both channels. 
The strength of the attraction is weaker than that in the $\Lambda N$ potential, which is consistent with phenomenological model calculations \cite{Dover, Bando}. 
For the tensor potential, we found that the strength is weaker compared with that for the $\Lambda N$ system. 

We next calculated phase shifts and scattering lengths using the potential fitted to the lattice data.
The results show that the $\Lambda_c N$ interaction is attractive at low energies ($E \lesssim 40$ MeV) in both $^1S_0$ and $^3S_1-^3D_1$ channels with comparable strength.
In order to quantify the similarity of potentials between two channels, we decomposed the central potentials into the spin-independent and spin-dependent ones. 
We found that the spin-dependent potential is negligibly weak except at short distances.

The dominant contribution of the $\Lambda_c N$ interaction comes from the spin-independent central potential, from which we have constructed a single-folding potential for $\Lambda_c$ hypernuclei.
We then estimated the binding energies of $\Lambda_c$ with the nuclei, $^{12}$C, $^{28}$Ni, $^{40}$Ca, $^{58}$Ni, $^{90}$Zr and $^{208}$Pb, by using the Gaussian expansion method. 
Resultant binding energies of $\Lambda_c$ hypernuclei become larger as the atomic number increases and/or the $u$, $d$ quark mass decreases. 
In order to estimate the effect of the Coulomb repulsion in the $\Lambda_c$ hypernuclei, we calculated expectation values of the folding potential of Coulomb force using the binding solution of $\Lambda_c$ hypernuclei.
The expectation value of the Coulomb potential is larger than the binding energy of $\Lambda_c$ hypernuclei for heavy nuclei, while that is comparable to the binding energies from QCD for the nuclei with $A=12-58$.
These suggest possible $\Lambda_c$ hypernuclei with light or medium-heavy nuclei in the real world. 

Currently, we plan to carry out full QCD simulations near the physical quark masses by using gauge configurations generated by K-computer in AICS, RIKEN. 
This may make it possible to draw definite conclusions on the $\Lambda_c N$ interactions and $\Lambda_c$ hypernuclei. 
We also plan to investigate the inelastic contributions for $\Lambda_c N$ interactions on the basis of the coupled-channel HAL QCD method \cite{HAL_CCP, Sasaki}. 
It is also an interesting future problem to study interactions in other two-body systems such as $\Lambda_c \Lambda_c$ and $\Xi_c N$, in order to understand the nature of charmed baryon interactions. 
\section*{Acknowledgments}
The author thanks all the members of the HAL QCD Collaboration for discussion. 
The author is also grateful to Shigehiro Yasui and Yasuhiro Yamaguchi for fruitful discussions and comments. 
This work is supported in part by the Grant-in-Aid of the Japanese Ministry of Education, Sciences and Technology, Sports and Culture (MEXT) for Scientific Research (No.~JP15K17667, JP16H03978, JP17K14287, (C)26400281), by a priority issue (Elucidation of the fundamental laws and evolution of the universe) to be tackled by using Post ``K" Computer, and by Joint Institute for Computational Fundamental Science (JICFuS). 
T.D. and T.H. are partially supported by RIKEN iTHES Project and iTHEMS Program.
S.G is supported by the Special Postdoctoral Researchers Program of RIKEN.
We thank the PACS-CS Collaboration for providing us their $2+1$ flavor gauge configurations \cite{PACS-CS}. 
Numerical computations of this work have been carried out by the KEK supercomputer system (BG/Q), [Project number : 14/15-21, 15/16-12].




\end{document}